\documentclass{article}

\usepackage{arxiv}

\usepackage[utf8]{inputenc} % allow utf-8 input
\usepackage[T1]{fontenc}    % use 8-bit T1 fonts
\usepackage{hyperref}       % hyperlinks
\usepackage{url}            % simple URL typesetting
\usepackage{booktabs}       % professional-quality tables
\usepackage{amsfonts}       % blackboard math symbols
\usepackage{nicefrac}       % compact symbols for 1/2, etc.
\usepackage{microtype}      % microtypography
\usepackage{graphicx}
\usepackage{amssymb}
\usepackage{amsmath}
\usepackage{lineno}
\usepackage{listings}
\usepackage{subfigure}
\usepackage{comment}
\usepackage{enumitem}
\usepackage[dvipsnames]{xcolor}

\title{The potential of self-supervised networks for random noise suppression in seismic data}

\author{
  Claire Birnie \\
  KAUST\\
  Thuwal, Kingdom of Saudi Arabia \\
  \texttt{claire.birnie@kaust.edu.sa}
   \And
  Matteo Ravasi \\
  KAUST\\
  Thuwal, Kingdom of Saudi Arabia \\
   \And
  Tariq Alkhalifah \\
  KAUST\\
  Thuwal, Kingdom of Saudi Arabia \\
   \And
  Sixiu Liu \\
  KAUST\\
  Thuwal, Kingdom of Saudi Arabia
}
\begin{document}

\chead{Seismic denoising by blind spot networks}

\maketitle

\begin{abstract}
  Noise suppression is an essential step in any seismic processing workflow. A portion of this noise, particularly in land datasets, presents itself as random noise. In recent years, neural networks have been successfully used to denoise seismic data in a supervised fashion. However, supervised learning always comes with the often unachievable requirement of having noisy-clean data pairs for training. Using blind-spot networks, we redefine the denoising task as a self-supervised procedure where the network uses the surrounding noisy samples to estimate the noise-free value of a central sample. Based on the assumption that noise is statistically independent between samples, the network struggles to predict the noise component of the sample due to its randomnicity, whilst the signal component is accurately predicted due to its spatio-temporal coherency. Illustrated on synthetic examples, the blind-spot network is shown to be an efficient denoiser of seismic data contaminated by random noise with minimal damage to the signal; therefore, providing improvements in both the image domain and down-the-line tasks, such as inversion. To conclude the study, the suggested approach is applied to field data and the results are compared with two commonly used random denoising techniques: FX-deconvolution and Curvelet transform. By demonstrating that blind-spot networks are an efficient suppressor of random noise, we believe this is just the beginning of utilising self-supervised learning in seismic applications.
\end{abstract}

\section{Introduction}
Noise consistently appears as an unwanted companion to the desired signal in seismic recordings. As such, noise suppression is a fundamental step in all seismic processing workflows \cite{Yilmaz2001}. Arising from local site conditions, as well as being excited by a seismic source, the total noise field can be seen as the sum of many noise components arising from different sources, each with their own characteristics \cite{Birnie2016}. Typically, noise suppression procedures identify a defining property that easily distinguishes the targeted noise from the desired signal and leverage that to separate the former from the latter. In this paper, we consider the random component of the noise field and leverage its non-predictable nature to build a suppression procedure.

Random noise suppression has been extensively investigated by the seismic community with the majority of the proposed techniques falling into one of the following categories: prediction-, transformation- and decomposition-based. Prediction-based approaches typically employ prediction-filters which aim to leverage the predictable nature of the coherent signal and therefore act as noise suppressors. Examples of such approaches include t-x predictive filtering and f-x deconvolution both of which can be applied in a stationary or non-stationary manner (e.g., \cite{chase1992,abma1995,gulunay2000,liu2013}). Transformation-based approaches transform the data into a domain in which the signal and noise can be easily distinguished due to their individual characteristics. By exploiting the sparse nature of seismic data in the curvelet domain, the curvelet transform is an example of a commonly used transformation-based denoising procedure  (e.g., \cite{hennenfent2006,neelamani2008,lianyu2009}). Similarly, other transformation-based methods have been proposed in the literature that use different transforms, for example, the wavelet- \cite{zhang2003}, shearlet- \cite{merouane2015}, and seislet-transforms \cite{fomel2010}, among others. Finally, decomposition-based procedures express the seismic data as the composition of weighted basis functions and suppress those associated to the noise components. Such decomposition procedures utilise the likes of spectral decomposition \cite{fomel2013}, Emperical Mode Decomposition \cite{bekara2009}, and Singular Value Decomposition \cite{bekara2007}, among others.  

With the increased interest in the use of Machine Learning (ML) in geophysics, a new class of random noise suppression procedures have been proposed. The majority of these approaches fall into the realm of Deep Learning (DL) and use a supervised training approach, which requires clean data for training to accompany the noisy input data. As is prevalent across seismic applications of DL, a number of studies have considered the use of synthetic seismic datasets for training a Convolutional Neural Network (CNN) (e.g., \cite{si2019,kim2019,wang2019}). Whilst these experiments have shown promising denoising results on synthetic datasets they struggle on field data \cite{zhang2019}. Alternatively, others have used conventional denoising procedures to create their `clean' counterparts to the noisy input data for their CNN denoising procedures (e.g., \cite{mandelli2019}). However, as no suppression procedure is perfect, noise residual is likely to exist within the clean counterparts and is likely to hamper the networks denoising capabilities. Unsupervised DL procedures have no such requirements of noisy-clean pairs for training. Recently, \cite{zhang2019} illustrated how an encoder-decoder network could be trained on noisy seismic data for random noise attenuation whilst \cite{qiu2021} detailed how an alternative convolutional network architecture can be used without any requirement on windowing the data. Both these approaches were shown to outperform the FX-deconvolution noise suppression procedure.

Considering the broader scientific community, most DL approaches for random noise suppression of images, or image-like data, are typically supervised and therefore have the requirement of paired noisy-clean datasets for training \cite{lehtinen2018}. This is often an unrealistic requirement - not just in seismology but across many other fields where there is no monitoring technique in which a clean dataset can be collected. In 2018, \cite{lehtinen2018} proposed Noise2Noise which illustrated how, under the assumption of a stationary signal, a Neural Network (NN) could be trained to denoise an image based on training over two noisy samples. Whilst this removes the requirement of noisy-clean pairs, it requires noisy-noisy pairs in which the signal is consistent but the noise is varying within each pair - a problematic requirement for many monitoring applications. Building on this, \cite{krull2019} proposed Noise2Void (N2V) which requires only a single noisy image for training. Under the assumption that noise is statistically independent between samples, a blind-spot network is used to predict a central sample's value based on neighbouring samples. As the noise is independent between samples, the noise's contribution to the sample's value cannot be predicted and therefore only the signal's contribution is predicted, resulting in an efficient denoising procedure. Whilst N2V is an ML approach, it can also be considered as a prediction-based approach; wherein, it leverages the ability to predict the signal and the inability to predict the noise resulting in a denoised image.

Previously applied to natural images and microscopy data among others, in this paper we investigate the adaptation of the N2V workflow to handle the highly oscillatory seismic signal and pseudo-random noise. Through an extensive hyper-parameter analysis, we identify the optimum hyper-parameters for the seismic denoising scenario considering both immediate improvements in the image domain and those observed for down-the-line tasks, such as seismic inversion. The paper concludes by illustrating the potential of N2V through an application to a field dataset.

\section{Theory: Blind-spot networks}

\begin{figure*}
  \centering
  \includegraphics[width=0.7\textwidth]{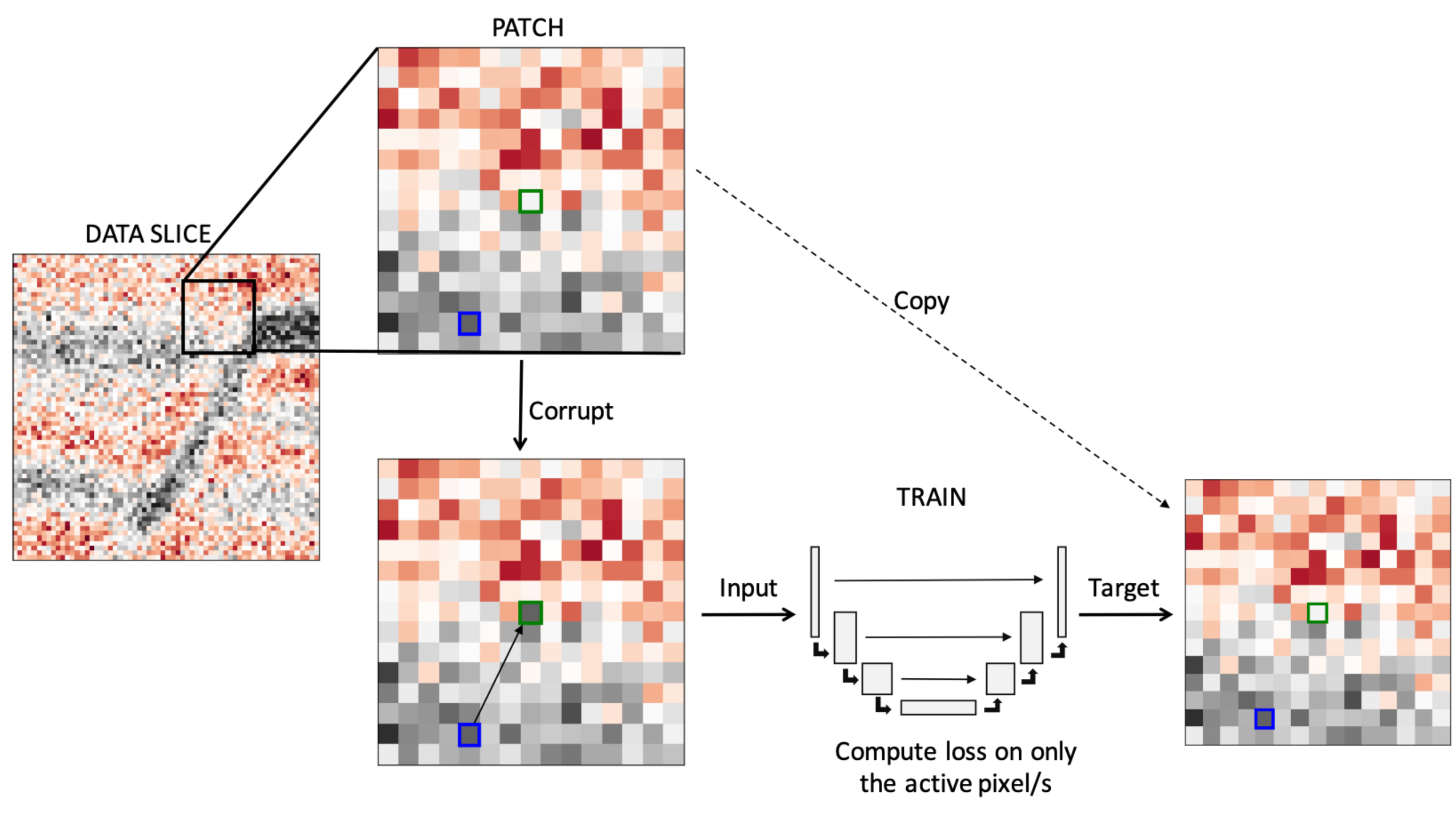}
  \caption{Schematic illustration of the Noise2Void denoising procedure.}
  \label{fig:workflow}
\end{figure*}

Noise2Void \cite{krull2019} is based on the concept of blind-spot NNs, which aim to predict a central pixel value based on neighbouring pixels. Operating on patches $x^j$ of a single image $x$, N2V works by replacing a set of non-adjacent pixels $x_i^j \; i=1,2,N_p$ from each patch, herein referred to as active pixels, with randomly selected neighbouring pixels, that pertain to the receptive field of the chosen network $\Omega_i^j$, as illustrated in Figure \ref{fig:workflow}. The corrupted patches become the input to a NN whilst the corresponding original patches represents the target values. In theory, the NN architecture, denoted as $f_\theta$ where $\theta$ refers to the trainable parameters, could be anything that can realistically map between the input and target values. In this paper, we follow the original N2V NN architecture: a 2-layer UNet styled after \cite{ronneberger2015}, as illustrated in Figure \ref{fig:workflow}. As opposed to standard NN image processing tasks, the loss function here is not computed on every pixel in the image, instead it is only evaluated for the active pixels, i.e., those that were corrupted in the input image:
%\begin{equation}
%\label{eq:aetraining}
%\hat{\theta} = \underset{\theta}  {\mathrm{argmin}} \frac{1}{N_s} %\sum_{i=1}^{N_s} ||\mathbb{I}_{\mathcal{J}}(f_\theta(\mathbb{I}_{\mathcal{J}' \rightarrow \mathcal{J}}(x_i))) - \mathbb{I}_{\mathcal{J}}(x_i)||_p
%\end{equation}
\begin{equation}
\label{eq:aetraining}
\hat{\theta} = \underset{\theta}  {\mathrm{argmin}} \frac{1}{N_s N_p} \sum_{j=1}^{N_s} \sum_{i=1}^{N_s} |x^j_i - f_\theta(\Omega^j_{x_i}) |^p
\end{equation}

where $p=1,2$ refers to the norm used in the loss - Mean Absolute Error (MAE) or Mean Squared Error (MSE), respectively - and $N_s$ is the number of available training samples (i.e., patches extracted from the image). \cite{krull2019} illustrated how MSE is the preferred choice for denoising additive WGN with respect to N2V. However, MAE is a prevalent choice for seismic deep learning applications. In Appendix 1, we provide a mathematical formulation that explains under which circumstances MSE and MAE should be used, respectively. Finally, we also highlight that no theoretical guarantee can be provided in the case of correlated noise therefore, we decided to experiment with both loss functions in the following numerical examples.

Once trained, the model is applied directly to the full seismic data. Note that at this stage, windowing of the seismic data is not required due to the ability of CNNs to handle dynamically varying input sizes. However, in the scenario where the data dimensions are not compatible with the down-/up-sampling of the UNet, then the input data are zero-padded to achieve an acceptable input data size.

\subsection{Performance metrics}
Whilst theoretically, N2V can be applied at any processing stage where random noise is observed in the seismic data, in this paper we focus on seismic images after time migration. A common challenge with many denoising procedures is that as part of the noise suppression process not only is the noise suppressed but also the signal is significantly damaged; something that may have a negative effect on down-the-line tasks. With this in mind, three performance metrics are considered: 
\begin{enumerate}[topsep=0pt,itemsep=-1ex,partopsep=1ex,parsep=1ex]
    \item Image Peak Signal-to-Noise Ratio (PSNR),
    \item Frequency correlation, and
    \item Post-stack inversion PSNR.
\end{enumerate}

The image PSNR is calculated as
\begin{equation}
    PSNR = 10 \cdot \log_{10}\left(\frac{max\{x\}^2}{n_x n_t||\hat{x}-x||^2}\right) ,
\label{eq:psnr}
\end{equation}
where $x$ represents the clean data, $\hat{x}$ the modified data (either noisy or denoised), and $n_t$ and $n_x$ represent the number of time samples and receivers, respectively. To quantify the effect of the N2V denoising, we compute the percent change in the PSNR between the noisy and denoised images. This can be written as,
\begin{equation}
    \%PSNR = \frac{100}{PSNR_{noisy}} \cdot PSNR_{N2V} ,
\end{equation}
where $PSNR_{noisy}$ and $PSNR_{n2v}$ are the $PSNR$ values computed from the noisy and denoised images, respectively.

The change in frequency is quantified using the sample Pearson's correlation coefficient, $r_{xy}$ where the aim is to return the noisy data's amplitude spectra to that of the clean data. This is computed as:
\begin{equation}
    r_{X,Y} = \frac{\sum_{i=0}^{n_f}(X_i-\Bar{X})\sum_{i=0}^{n_f}(Y_i-\Bar{Y})}
    {\sqrt{\sum_{i=0}^{n_f}(X_i-\Bar{X})^2}\sqrt{\sum_{i=0}^{n_f}(Y_i-\Bar{Y})^2}
    } ,
\label{eq:pearson}
\end{equation}
where $X$ and $Y$ are the amplitude spectra of the clean and denoised data, respectively, averaged over the spatial-axis, $\Bar{X}$ and $\Bar{Y}$ are the sample means of $X$ and $Y$, respectively, and $n_f$ is the number of samples in the spectra. Similarly to the image PSNR, to analyse the effect of N2V we compute the percent change in the sample Pearson's correlation coefficient when computed with the noisy versus denoised data, 
\begin{equation}
    \%r = \frac{100}{r_{cl,n}} \cdot r_{cl,N2V} ,
\end{equation}
where $r_{cl,n}$ is the correlation coefficient between the clean and noisy data and $r_{cl,N2V}$ is the correlation coefficient between the clean and denoised data.

As a final metric of comparison, the denoised data are used as input for a standard down-the-line task, namely post-stack inversion \cite{veeken2004}. By doing so, we can inspect the effect of denoising and possible signal damage on our ability to estimate an acoustic impedance model of the subsurface. Post-stack inversion assumes that each trace of the post-stack seismic data can be represented by a simple convolutional model as follows:
\begin{equation}
d(t) = \frac{1}{2} w(t) * \frac{d\ln(AI(t))}{\quad dt} ,
\label{eq:convmodel}
\end{equation}
where $AI(t)$ is the acoustic impedance profile in the time domain and $w(t)$ is the time domain seismic wavelet. We rewrite this expression for the entire set of traces in the post-stack data in a compact matrix-vector notation, $\mathbf{d}= \mathbf{W} \mathbf{D} \mathbf{m}$, where $\textbf{d}$ and $\textbf{m}$ are vectorized seismic data and the natural logarithm of the acoustic impedance model, $\mathbf{W}$ is a convolution operator and $\mathbf{D}$ is a first derivative operator. The model vector is then estimated by means of L2-regularized inversion using the PyLops computation framework \cite{Ravasi2020}. The PSNR values for the inverted models are calculated as in equation \ref{eq:pearson} where $x$ is the true model and $\hat{x}$ is the inverted model. As with the above two performance metrics, for the inversion PSNR we calculate the percent change between the inversions for the noisy and denoised inversions.

Finally, for the field dataset there is no `clean' dataset for comparison and as such, the above metrics cannot be easily computed. In this situation, we perform qualitative comparison of the raw and denoised images, frequency content, and inversion products. In addition to this, we benchmark the N2V approach against two commonly used techniques for random noise suppression: FX-deconvolution and sparsity-promoting inversion with Curvelet transform. FX-deconvolution \cite{Canales1984} is based on the concept that the coherent part of a seismic trace can be predicted in the FX domain from the previous traces via spatial predictive filters. Our results are based on the Madagascar implementation of FX-deconvolution \cite{gulunay1986} with the same parameters used by \cite{liu2018} for this dataset, namely a window of 20 traces with a filter length of 4 traces. The second method is instead based on the principle that seismic data have a sparse representation in the Curvelet domain \cite{neelamani2008} whilst random noise maps incoherently across the Curvelet domain. We employ sparsity-promoting inversion with the Fast Iterative Shrinkage-Thresholding Algorithm (FISTA) solver \cite{beck2009} and soft-thresholding to attenuate random noise in our seismic data. 

\section{Data}
\begin{figure*}[ht!]
  \centering
  \includegraphics[width=0.8\textwidth]{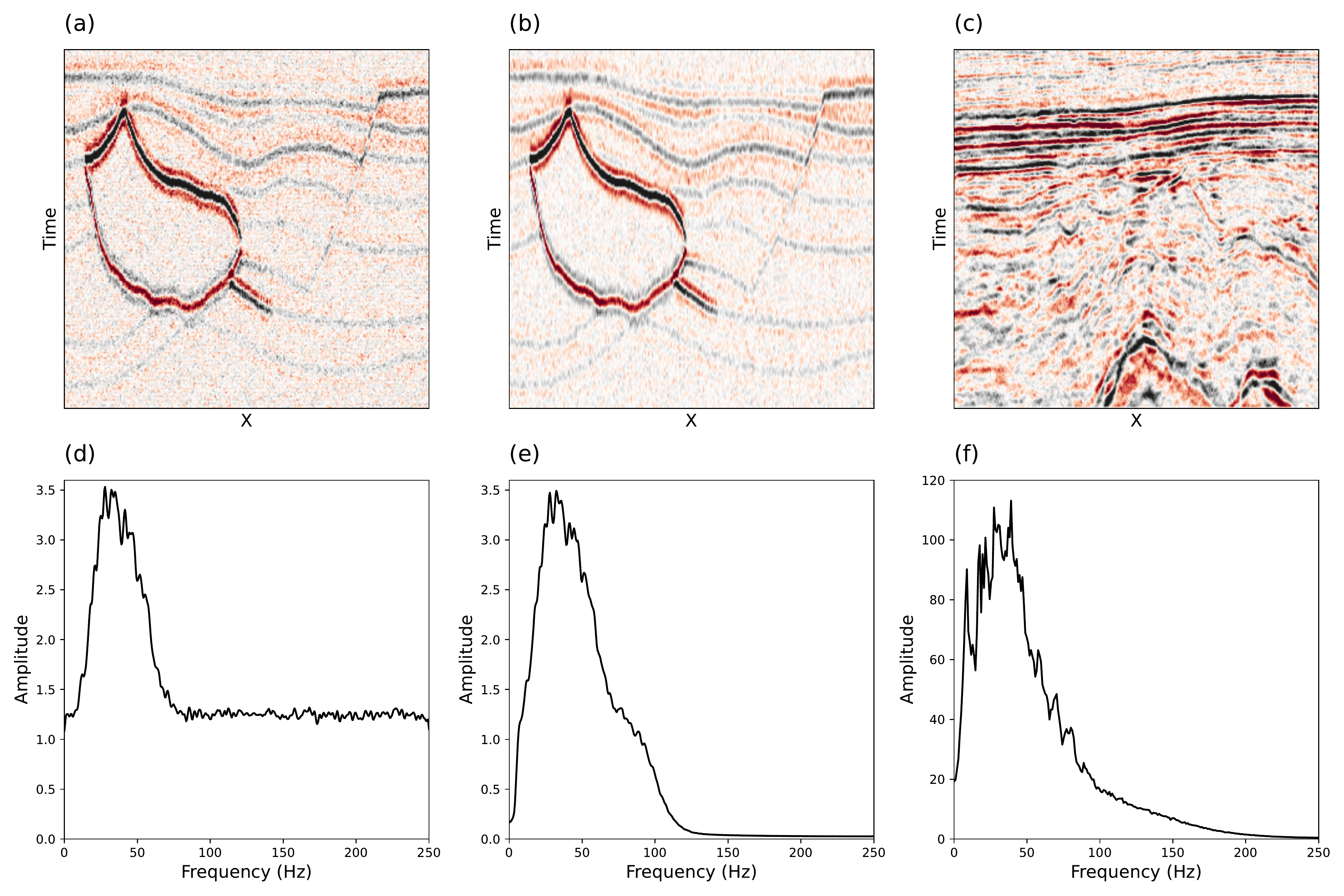}
  \caption{Datasets (top) used for training and testing N2V and their respective amplitude spectra (bottom): (a,d) Hess synthetic dataset with WGN, (b,e) Hess synthetic dataset with 5-100 Hz band-passed noise, and (c,f) land field dataset.}
  \label{fig:datasets}
\end{figure*}

To be able to perform a quantitative analysis of our denoising procedure the majority of this study is performed on synthetically generated data. Utilising the SEG Hess VTI model and a 30 Hz Ricker wavelet, $30$ 2D slices are created in order to mimic the multiple in-lines or cross-lines that would be available from a 3D survey. Two different synthetic datasets are created with this base waveform data: one with White, Gaussian Noise (WGN) as illustrated in Figure \ref{fig:datasets}(a,d) and one with 5-100 Hz band-passed noise as illustrated in Figure \ref{fig:datasets}(b,e).

The paper concludes with an application of the N2V denoising procedure on a field dataset from a land acquisition in China that is heavily contaminated by random noise. Previously analysed by \cite{liu2013} and \cite{liu2018}, a single 2D line from the post-stack volume was released under the Madagascar framework as part of a continued effort towards reproducible research. Figure \ref{fig:datasets}(c,f) illustrates the seismic image and its respective amplitude spectra. The sampling rate for the datasets (synthetic and field) is 2~ms.

\subsection{Data preparation}
\begin{figure*}[!htb]
  \centering
  \includegraphics[width=0.8\textwidth]{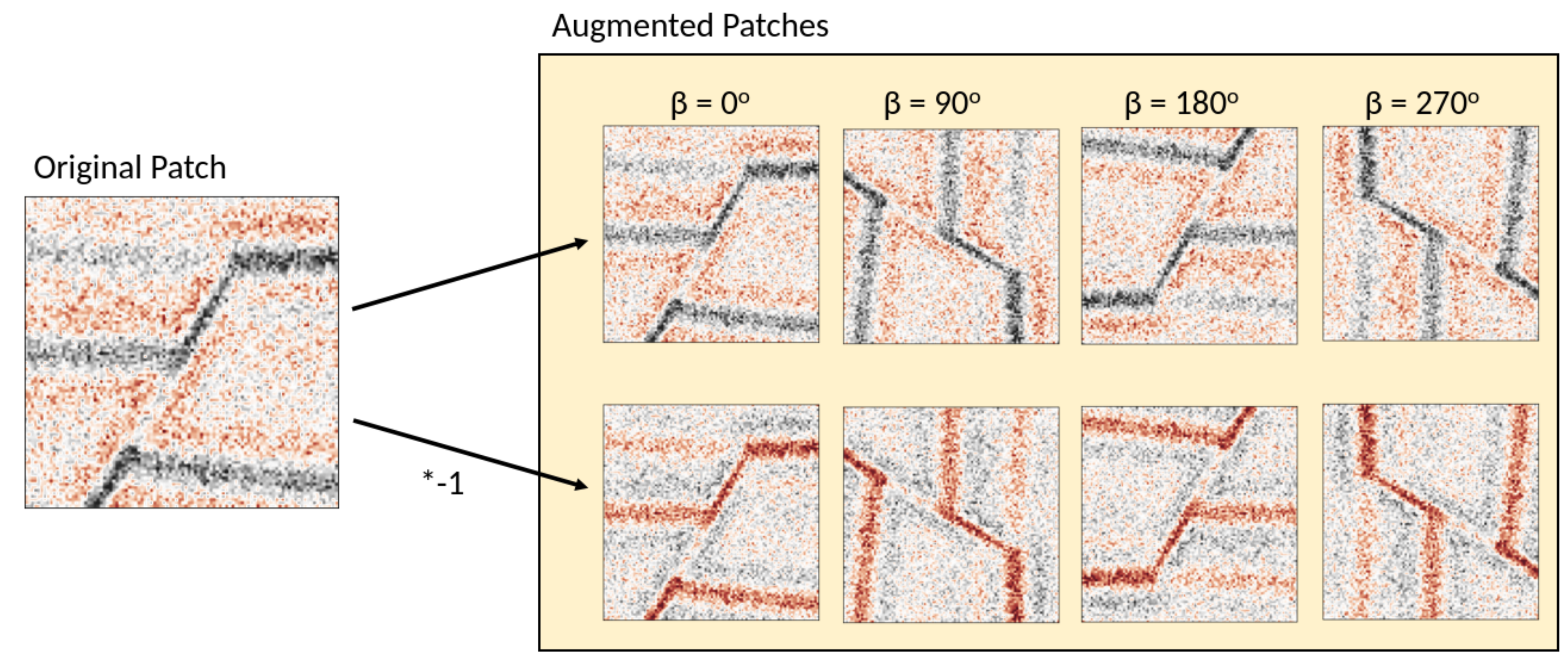}
  \caption{Data augmentation techniques of polarity reversal and rotation ($\beta$) applied to increase the size of the training data. }
  \label{fig:augmentation}
\end{figure*}

Following the procedure of N2V, patches are randomly extracted from the seismic lines to form the training dataset. To further increase the size of such dataset, common data augmentation techniques of both rotation and polarity reversal are employed. This results in the data increasing 8-fold as illustrated in Figure \ref{fig:augmentation}. The number and size of the training patches varies between the examples in this paper and are detailed in Table \ref{tab:hyperparams}.

\begin{table*}[!htb]
\centering
\caption{Hyper-parameters of N2V method for the three examples presented in this paper. Those chosen for Exp.2: BP Noise are selected from a hyper-parameter sweep, whilst those for Exp.3: Land Data are manually tuned initialised from those of Exp.2: BP Noise.}
\label{tab:hyperparams}
\begin{tabular}{|l|c|c|c|}
\hline
 & \multicolumn{1}{l|}{\textbf{Exp.1: WGN}} & \multicolumn{1}{l|}{\textbf{Exp.2: BP Noise}} & \multicolumn{1}{l|}{\textbf{Exp.3: Land Data}} \\ \hline
\textbf{Train-Validate} & 4500-500 & 4500-500 & 4500-500 \\ \hline
\textbf{Epochs} & 150 & 25 & 15 \\ \hline
\textbf{Batch size} & 128 & 64 & 128 \\ \hline
\textbf{Patch size} & 64x64 & 32x32 &  32x32 \\ \hline
\textbf{\% Active pixels} & 0.2 & 25 & 33 \\ \hline
\textbf{Neighborhood radius} & 5 & 15 & 15 \\ \hline
\textbf{Loss} & MSE & MAE & MAE  \\ \hline
\textbf{UNet depth} & 2 & 2 & 2 \\ \hline
\textbf{Kernel size} & 3x3 & 3x3 & 3x3 \\ \hline
\end{tabular}
\end{table*}

\section{Numerical Examples}

\subsection{Synthetic with WGN}
\begin{figure}
  \centering
  \includegraphics[width=0.5\textwidth]{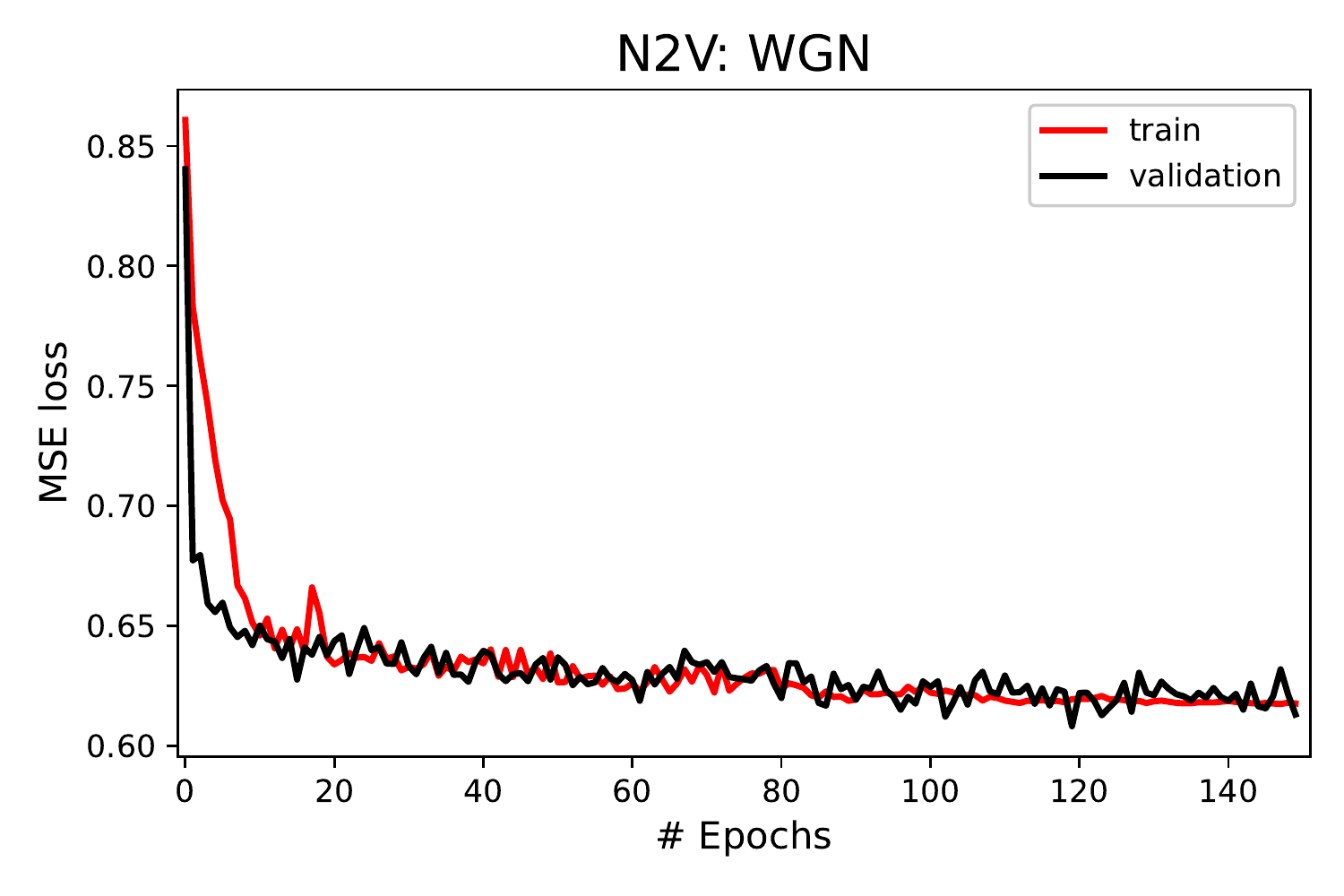}
  \caption{Progression of train (red) and validation (black) losses for N2V training on the Hess VTI model with additive WGN.}
  \label{fig:wgn_losses}
\end{figure}

\begin{figure*}
  \centering
  \includegraphics[width=0.85\textwidth]{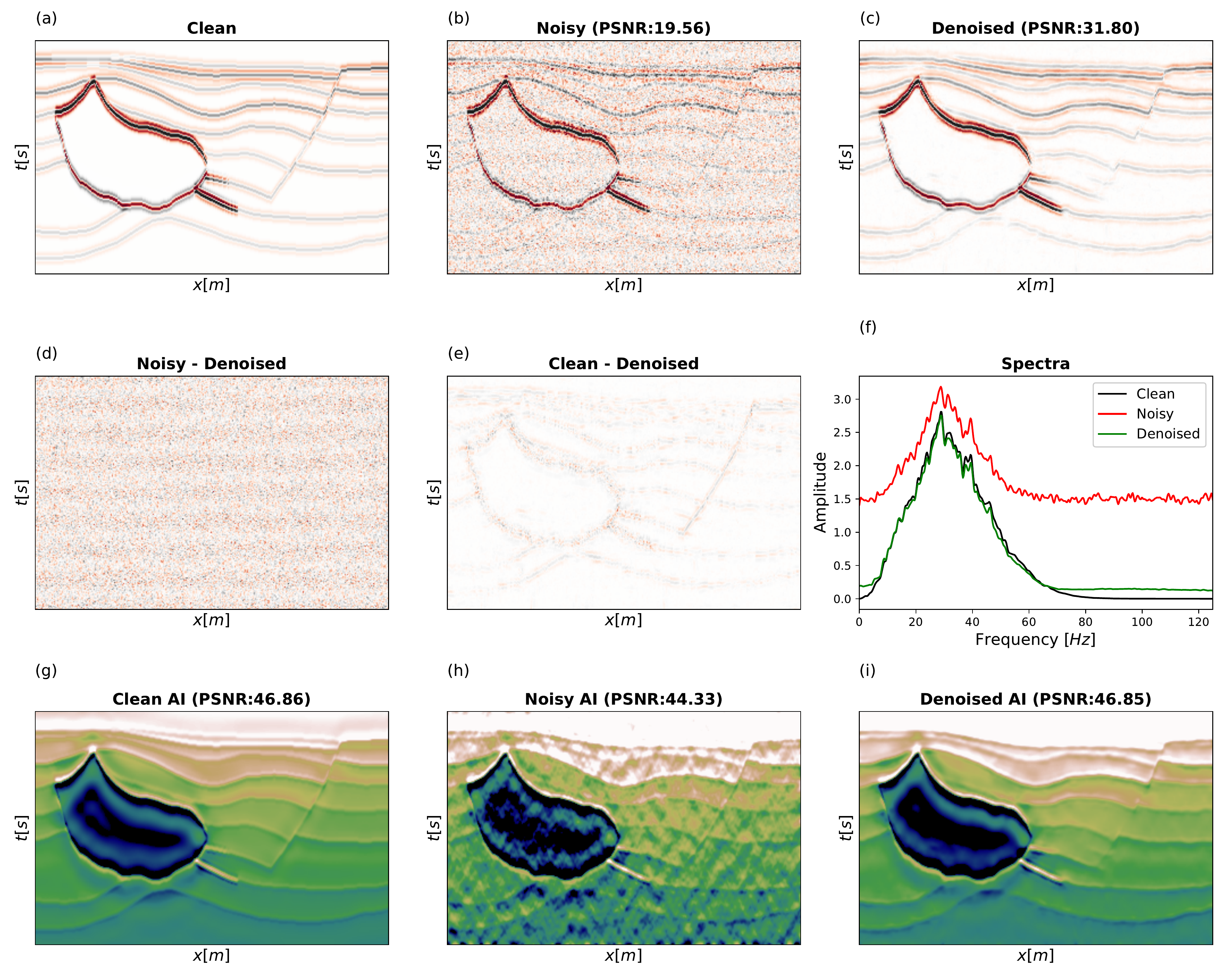}
  \caption{Trained N2V model applied to a synthetic dataset with WGN. (a) Noise-free synthetic, (b) noisy synthetic given as input to the model, and (c) result of the N2V denoising procedure. (d) and (e) portray the differences between the noisy and denoised datasets and between the noise-free and denoised datasets, respectively. Whilst (g), (h) and (i) are the results of an L2-regularised inversion for the clean, noisy and denoised data, respectively.}
  \label{fig:n2v_wgn}
\end{figure*}

The initial example portrays a layman’s application of N2V onto seismic data utilising the same noise properties (i.e., WGN) and hyper-parameters as detailed in the original N2V study on natural images \cite{krull2019} and displayed in Table \ref{tab:hyperparams}. Figure \ref{fig:wgn_losses} shows the progression of the loss (equation \ref{eq:aetraining}) during the training period for all the active pixels in the 4500 training patches and the 500 validation patches. An application of the trained model to a 2D line from the synthetic dataset is illustrated in Figure \ref{fig:n2v_wgn}. The training took $~12.5$ minutes on an Nvidia Quadro RTX 4000 while the application on the 2D line composed of 198 traces of 453 time samples took $~38$ milliseconds. In the image domain, the PSNR has increased by 73\% whilst in the frequency domain there is a much higher similarity between the spectra of the noise-free wavefield (black) and the denoised data (green), as opposed to the noisy data (red). When inversion is performed on the data, it is clear that the inversion on the denoised data produces an acoustic impedance model that is closer to the true model than that of the noisy data, with significantly fewer artefacts. Overall, for seismic data contaminated by WGN, it is shown that N2V can accurately learn to reproduce the seismic signal from surrounding samples without recreating the noise. This significantly improves the data quality both for current tasks (in the image domain) and down-the-line tasks, such as inversion. 

\begin{figure*}
  \centering
  \includegraphics[width=0.9\textwidth]{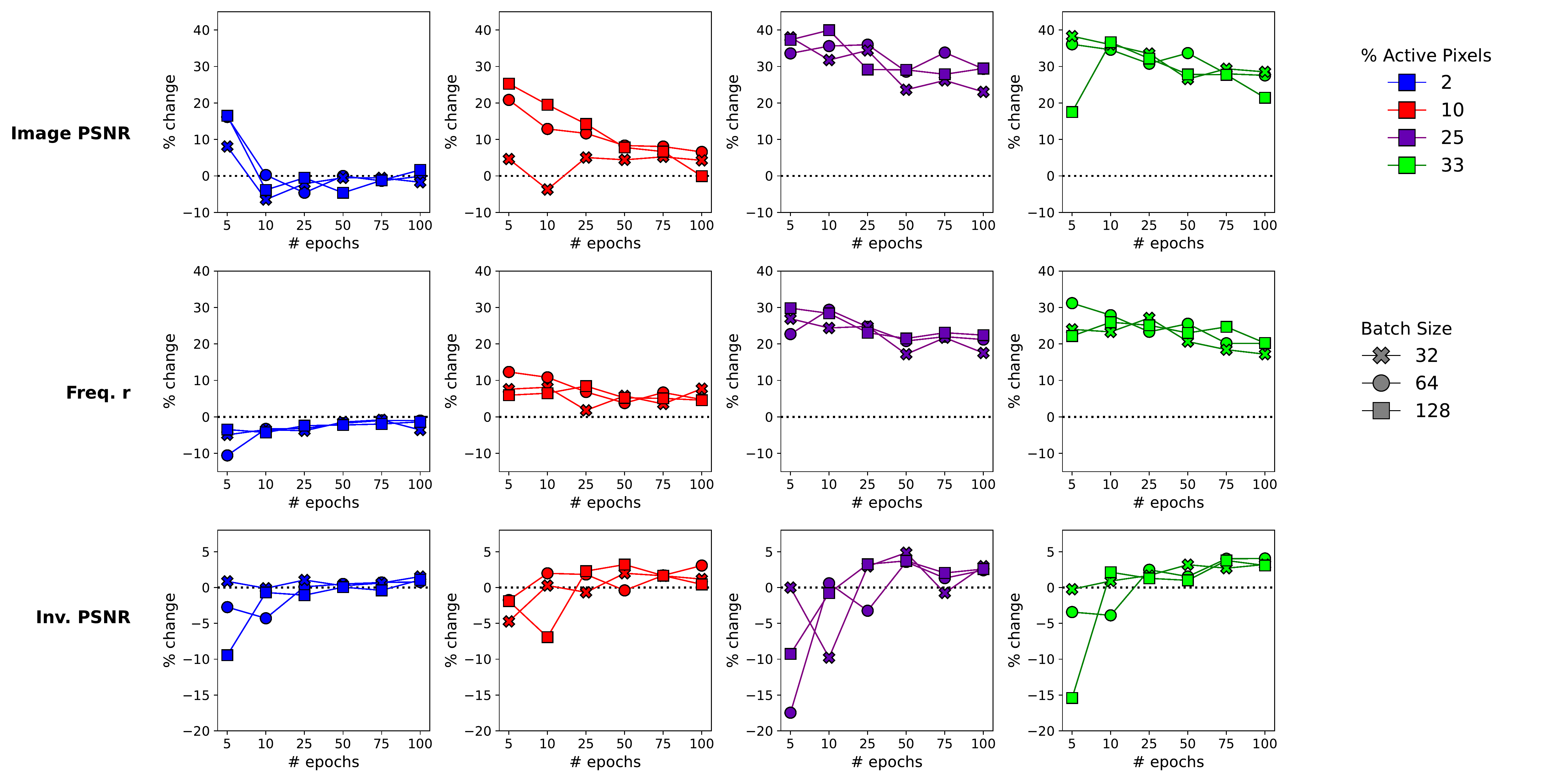}
  \caption{Subset of grid-search results on hyper-parameter selection of N2V for synthetic data with 5-100 Hz band-passed noise, a patch size of $32x32$, and the MAE loss function. The top row illustrates the \% change in the image PSNR, the middle row the \% change in the frequency correlation with the noise-free data, and the bottom row shows the \% change in the PSNR of the inverted acoustic impedance models. The colours represent the different number of active pixels per patch whilst the marker shapes represent the training batch size.}
  \label{fig:gridsearch}
\end{figure*}

\subsection{Synthetic with band-passed noise}
The second example focuses on providing a more realistic example using band-pass filtered noise to identify the potential of N2V for seismic applications. In this example, we have performed a hyper-parameter sweep to identify the optimum hyper-parameters for denoising of band-pass filtered random noise. Over 860 parameter combinations were considered from the values detailed in Table \ref{tab:gridsearch}. Due to the computational cost of training, only one model is generated per hyper-parameter combination however, 100 additional synthetic datasets are generated to analyse each models' performance. Figure \ref{fig:gridsearch} illustrates the performance of a subset of the hyper-parameters for a fixed window-size ($32-by-32$) and fixed loss function ($MAE$). 

\begin{table}[!htb]
\centering
\caption{Hyper-parameters considered during the hyper-parameter sweep aimed at identifying an optimal combination for denoising of 5-100 Hz band-passed noise.}
\label{tab:gridsearch}
\begin{tabular}{|l|c|}
\hline
 & \multicolumn{1}{l|}{\textbf{Parameter value options}} \\ \hline
\textbf{Epochs} & {[}5, 10, 25, 50, 75, 100, 150{]} \\ \hline
\textbf{Batch size} & {[}32, 64, 128{]} \\ \hline
\textbf{Patch size} & {[}32x32, 64x64{]} \\ \hline
\textbf{\% Active pixels} & {[}2, 10, 25, 33{]} \\ \hline
\textbf{Neighborhood radius} & {[}5, 15,30{]} \\ \hline
\textbf{Loss} & {[}MSE, MAE{]} \\ \hline
\textbf{Train-Validate} & 4500-500 \\ \hline
\textbf{UNet depth} & 2 \\ \hline
\textbf{Kernel size} & 3x3 \\ \hline
\end{tabular}
\end{table}

As detailed above, the models are evaluated on the PSNR gain in both the image and acoustic impedance domains, as well as the increase in the correlation with the amplitude spectra of the noise-free data. Considering  Figure \ref{fig:gridsearch}, the general trend of the experiments shows that as training progresses, i.e. the number of epochs increases, the PSNR gain in the image domain (top row) slightly decreases whilst the PSNR gain in the acoustic impedance domain (bottom row) moderately increases. Finally, the batch size is shown to have a limited influence in comparison to the other hyper-parameters. The optimum hyper-parameter combination is a sum of the ranking of each combination across all three scoring criteria.

\begin{figure}[!htb]
  \centering
  \includegraphics[width=0.5\textwidth]{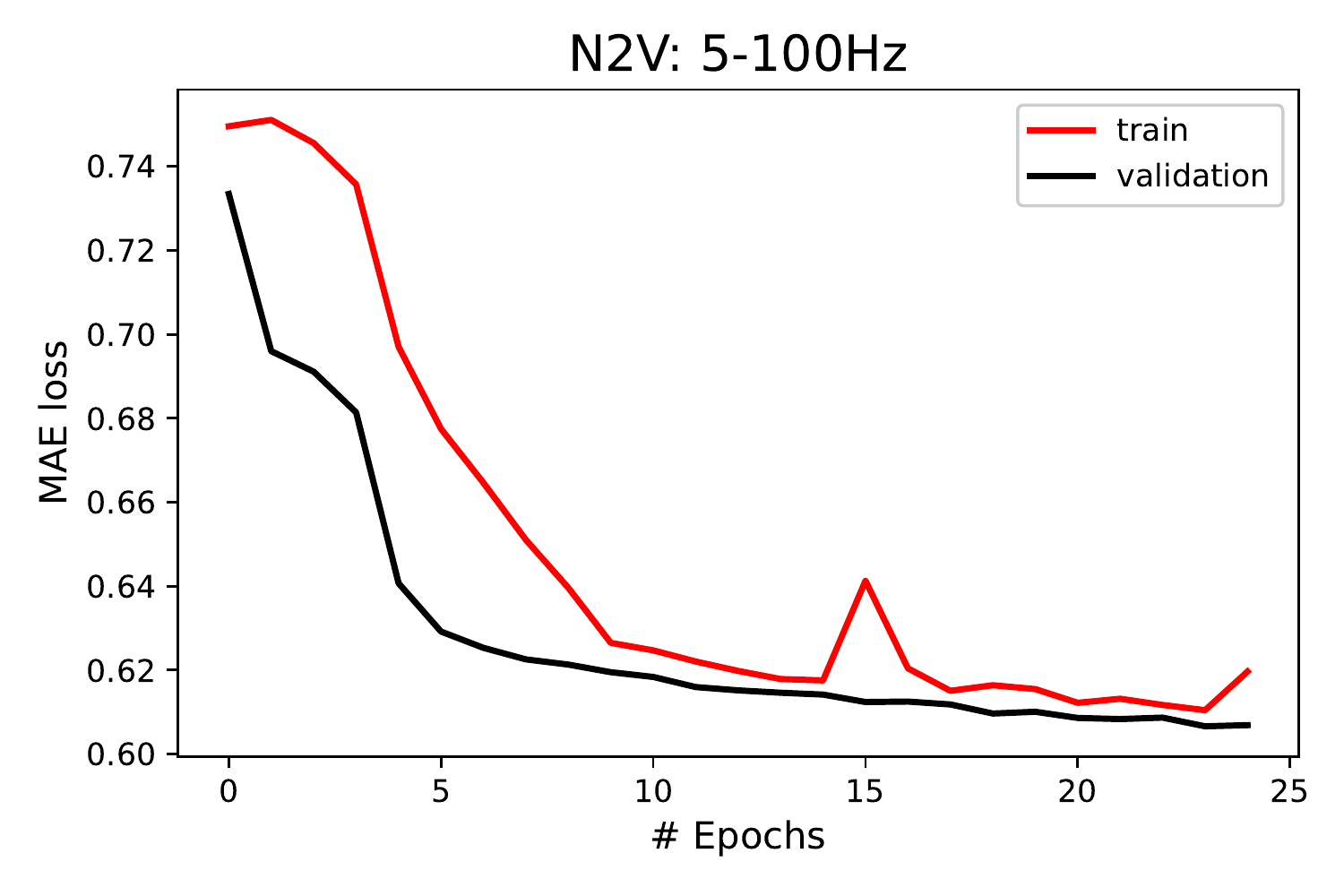}
  \caption{Progression of train (red) and validation (black) losses for N2V training on the Hess VTI model with additive $5-100$ Hz band-pass filtered noise.}
  \label{fig:bp_losses}
\end{figure}

\begin{figure*}[!htb]
  \centering
  \includegraphics[width=0.85\textwidth]{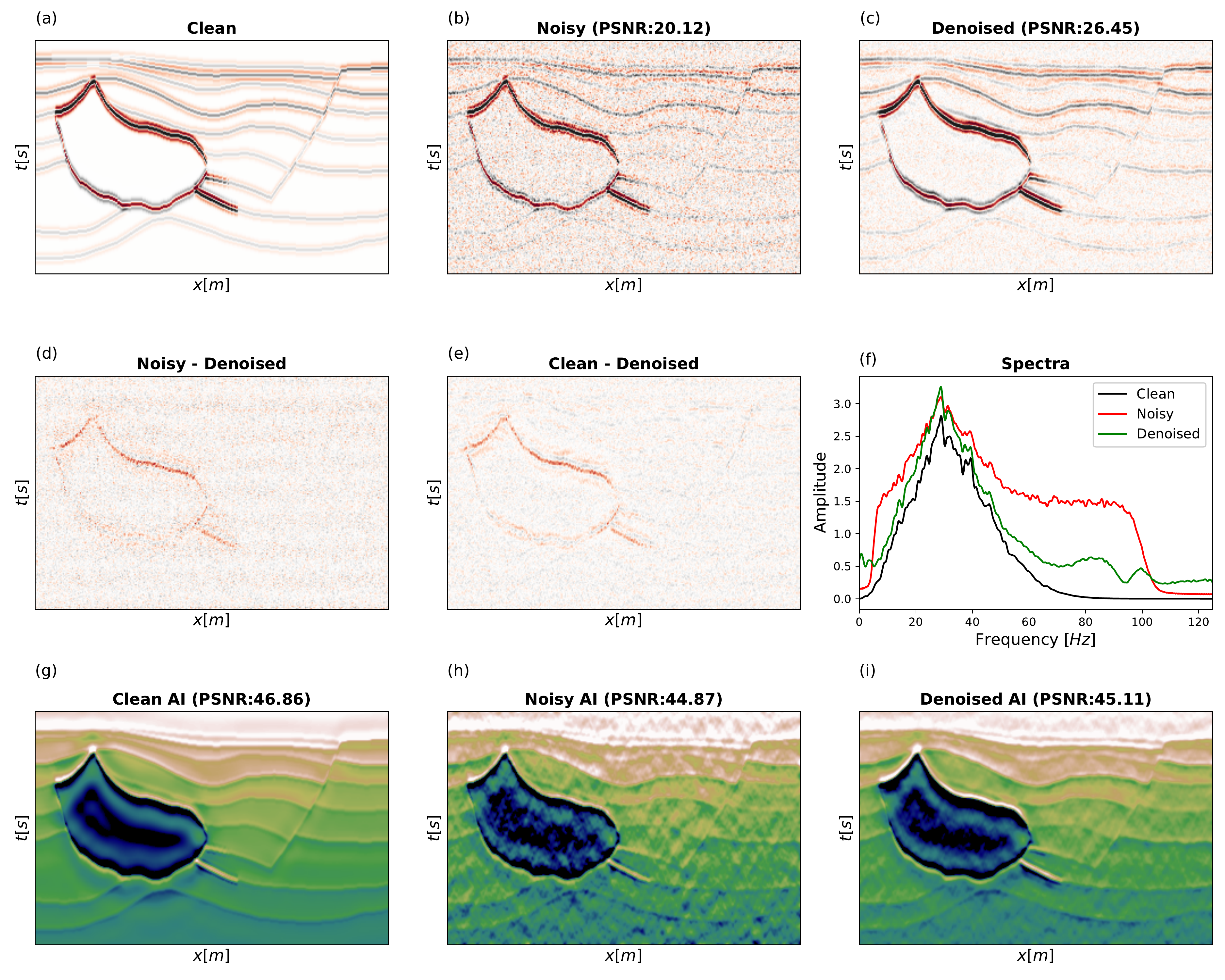}
  \caption{Trained N2V model applied to a synthetic dataset with $5-100$ Hz band-passed noise. (a) The noise-free synthetic, (b) the noisy synthetic given as input to the model, and (c) the result of the N2V denoising procedure. (d) and (e) portray the differences between the noisy and denoised datasets and between the noise-free and denoised datasets, respectively. Whilst (g), (h) and (i) are the results of an L2-regularised inversion for the clean, noisy and denoised data, respectively.}
  \label{fig:n2v_bp}
\end{figure*}

The optimum hyper-parameter combination, as detailed in the middle column of Table \ref{tab:hyperparams}, is used to train the band-passed N2V model with Figure \ref{fig:bp_losses} illustrating the loss function progression during training. Figure \ref{fig:n2v_bp} shows the result of the trained network applied to a synthetic slice contaminated by band-pass filtered noise. Similar to the WGN results, a PSNR increase is observed in the image domain, alongside an increase in the similarity with the amplitude spectra of the noise free data. However, more substantial signal leakage is observed around the central salt body in comparison to the WGN results, as well as some noise residual remaining in the denoised result. Despite this, the inversion on the denoised image results in a cleaner subsurface model than that from the noisy image.

\subsection{Field data application}
\begin{figure*}[!htb]
  \centering
  \includegraphics[width=1.\textwidth]{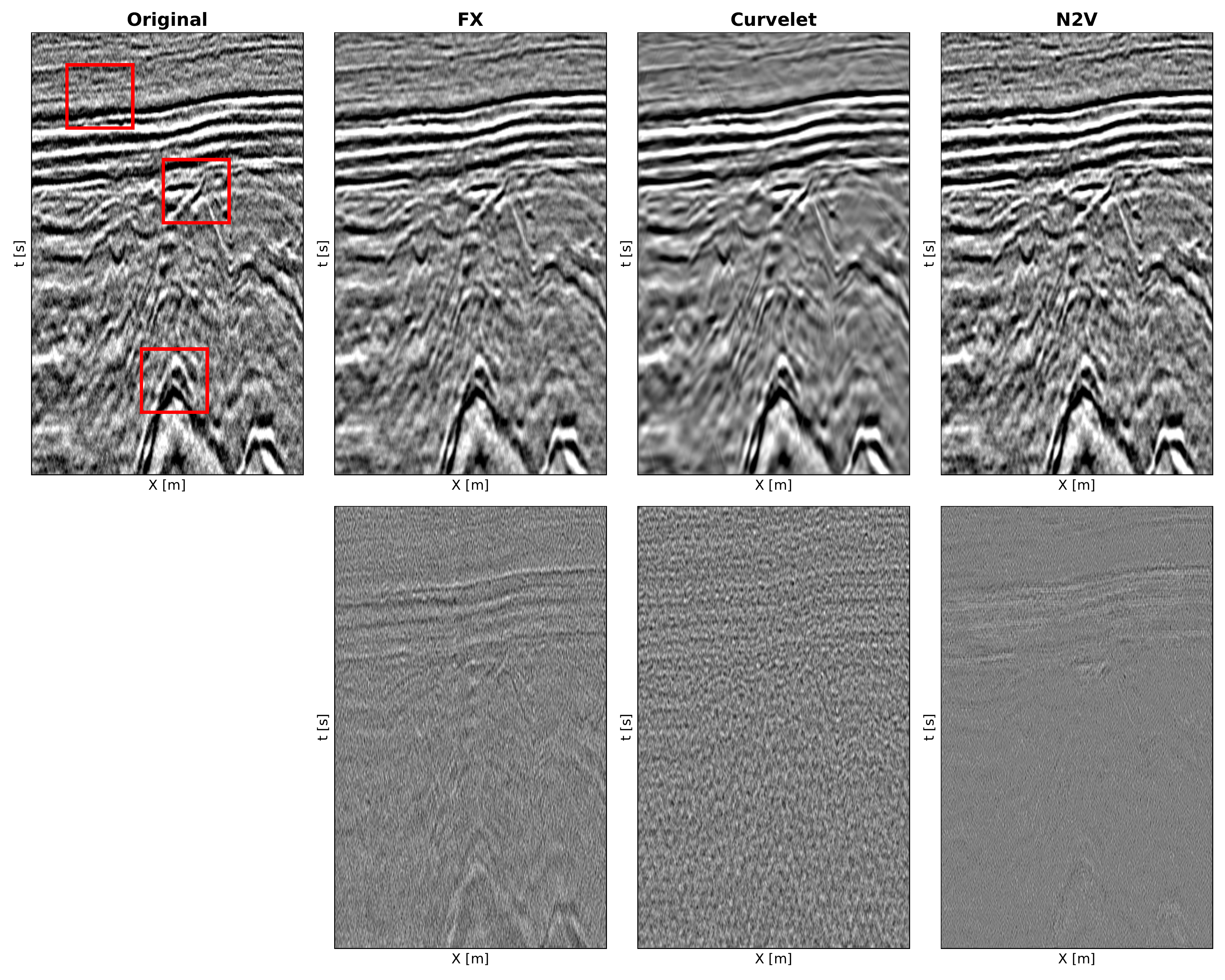}
  \caption{Comparison of different random noise suppression procedures. The top row shows the original data (left) followed by the results from the FX-, curvelet- and N2V-denoising procedures, from left to right. The bottom row illustrates the difference between the denoising results and the original data.}
  \label{fig:n2v_field_image}
\end{figure*}

\begin{figure*}
  \centering
  \includegraphics[width=1.\textwidth]{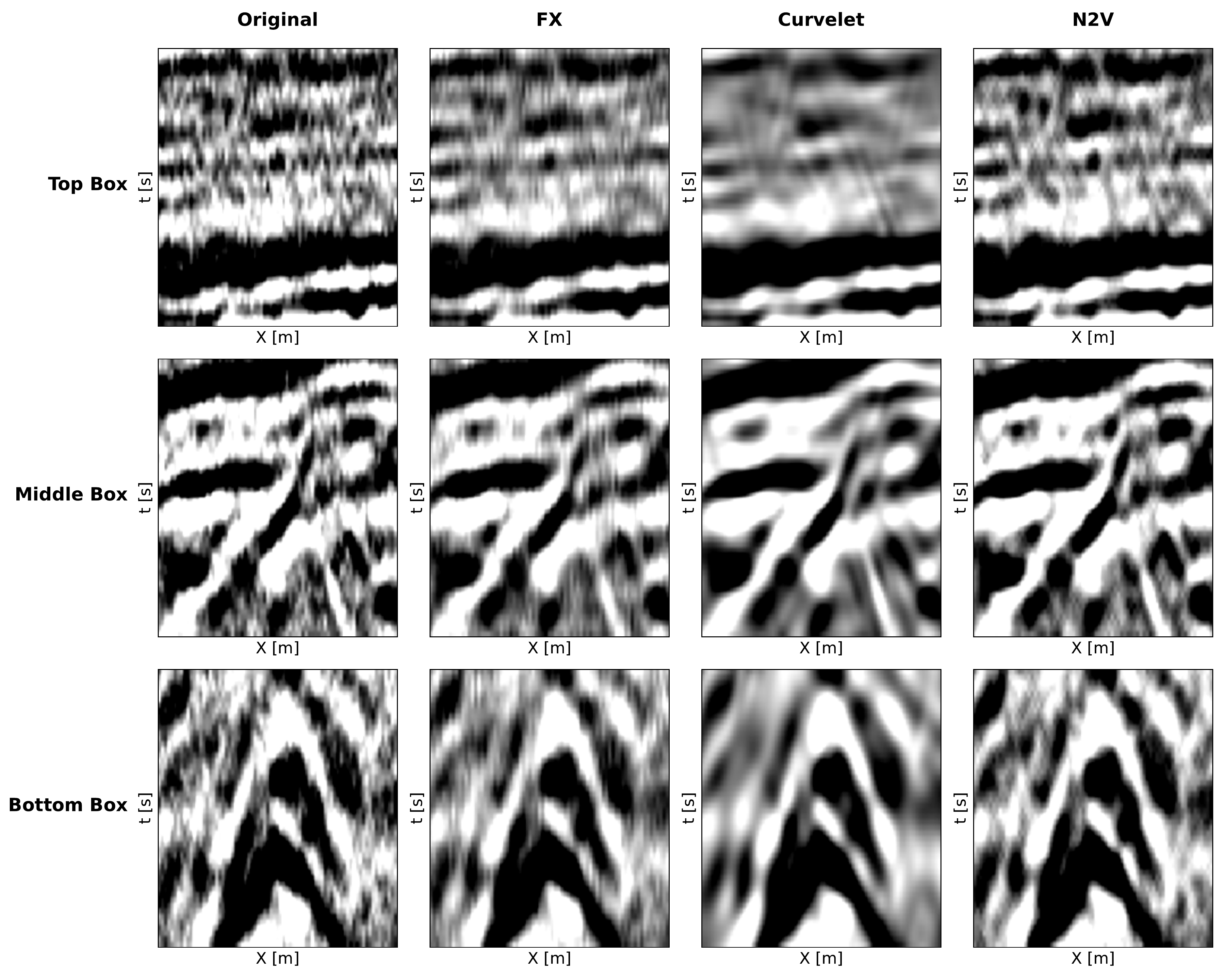}
  \caption{Close-up comparison of different random noise suppression procedures from areas highlighted in Figure \ref{fig:n2v_field_image}}
  \label{fig:n2v_field_image_closeup}
\end{figure*}

To conclude, the N2V workflow is applied to a land dataset, which is known to be contaminated by random noise. Trained using $4500$ patches, the model is applied to the full 2D line. The resulting denoised image is shown in Figure \ref{fig:n2v_field_image} and we compare it with results from the conventional FX-deconvolution and Curvelet denoising procedures. Figure \ref{fig:n2v_field_image_closeup} provides a zoomed-in comparison for three areas of interest spanning the model's depth range. Considering the difference between the original and denoised datasets (bottom row of Figure \ref{fig:n2v_field_image}), the Curvelet approach has removed the most noise however we argue that it has possibly over-smoothed the data  (effectively reducing the resolution) as well as introduced some linear artefacts (particularly noticable in the top row of Figure \ref{fig:n2v_field_image_closeup}). This is likely due to the fact that since the Curvelet transform explains an image as the superposition of localised oriented wave packets, the denoising process may have slightly corrupted the relative weighting of the different wave packets whilst trying to suppress the noise. On the other hand, while the FX approach has also removed more energy than the N2V approach, the resulting denoised image is still heavily contaminated by noise, which is particularly observable in the closeups of Figure \ref{fig:n2v_field_image_closeup}. In addition to this, all approaches have resulted in a certain amount of signal leakage, even though of different nature from method to method.

\begin{figure*}
  \centering
  \includegraphics[width=0.75\textwidth]{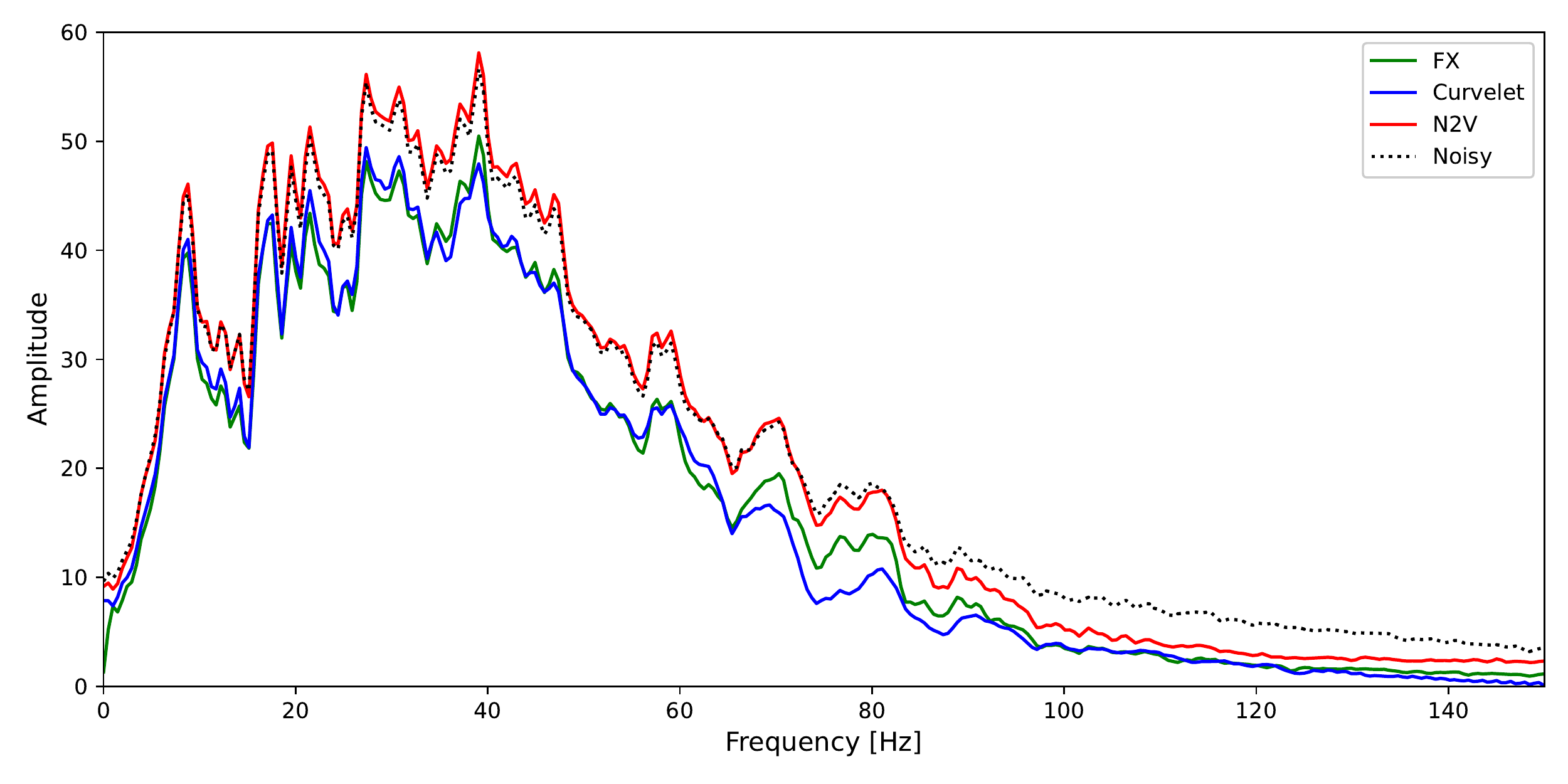}
  \caption{Comparison of the effect of the different random noise suppression procedures on the amplitude spectra of the data. }
  \label{fig:n2v_field_spectra}
\end{figure*}

Supporting what is observed in the image domain, Figure \ref{fig:n2v_field_spectra} illustrates the differences in the amplitude spectra between the different denoised datasets. Both the FX-deconvolution and Curvelet domain results have reduced the energy across all bandwidths with the Curvelet approach outperforming the FX approach between 60-100 Hz. The N2V results show less reduction in the bandwidths around which the signal is expected with a reduction in energy being observed above $~85$ Hz. However, even at higher frequencies where signal is likely not to be present, the N2V approach does not reduce the amplitude spectrum to the levels of the other two procedures. 

\begin{figure*}
  \centering
  \includegraphics[width=0.75\textwidth]{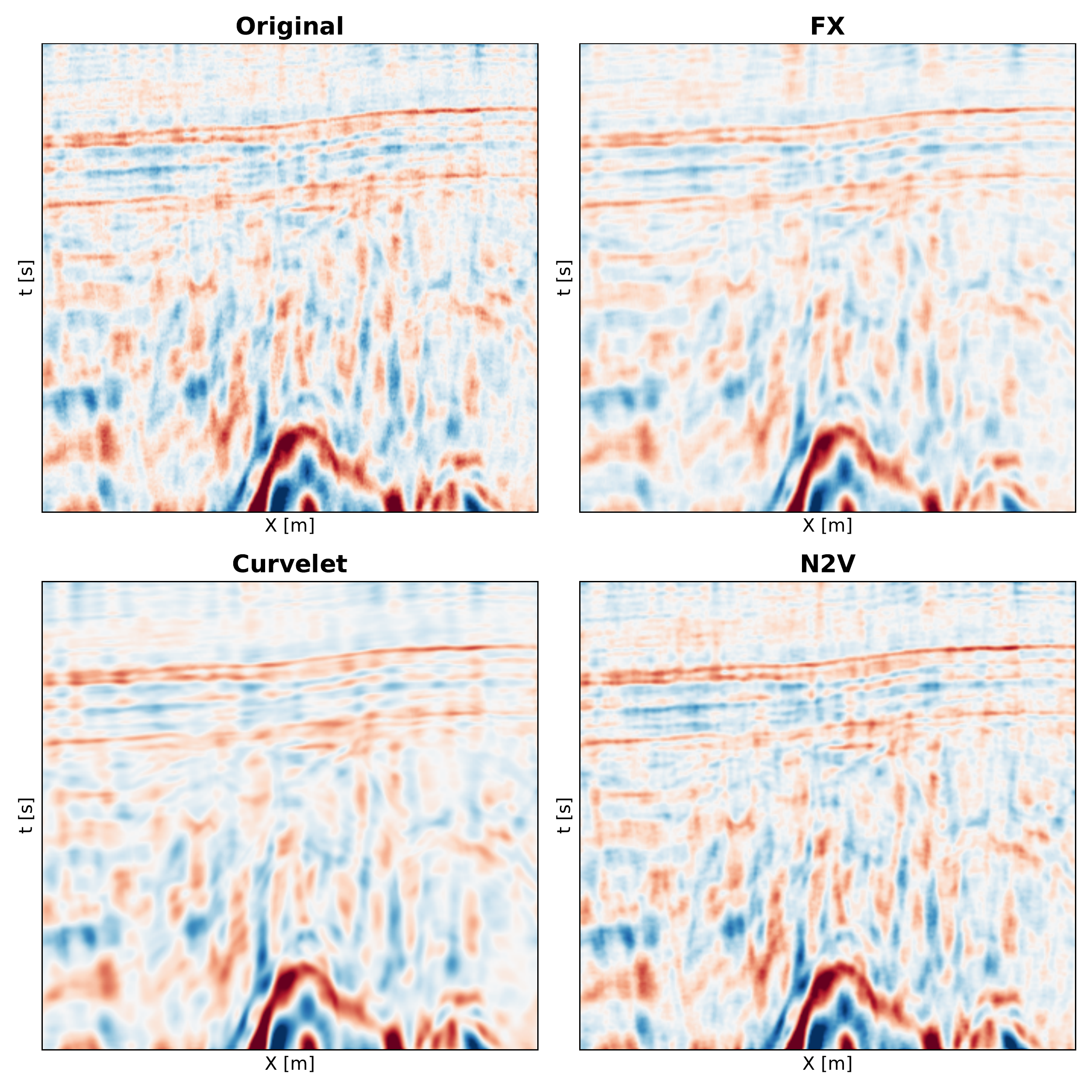}
  \caption{Comparison of the effect of the different random noise suppression procedures when the denoised datasets are fed into an L2 inversion. }
  \label{fig:n2v_field_inversion}
\end{figure*}

Finally, Figure \ref{fig:n2v_field_inversion} shows the inversion products for the original data and the denoised results. Similar to the observations in the image domain, the N2V results seem to have more details than the overly-smoothed Curvelet results. Conversely, the partial attenuation of genuine signal in the FX data leads to lower contrasts between features in the inverted acoustic impedance model.

\section{Discussion}
Blind-spot networks offer a solution to the predicament of requiring noisy-clean pairs of training images for deep learning denoising procedures. Previously utilised in applications on the likes of natural images \cite{krull2019}, Computed Tomography (CT) images \cite{liang2021} and Synthetic Aperture Radar (SAR) images \cite{molini2021}, we have shown that under the right circumstances, N2V can also be a powerful denoiser for seismic data. N2V relies on the assumption that noise is statistically independent between pixels, or, as in the seismic case, between each spatio-temporal sample. In reality, noise in seismic data is always correlated to some extent. Despite this, we have shown that, whilst providing the best results on WGN, N2V can still efficiently denoise both synthetic band-pass filtered noise as well as recorded noise from a field acquisition. It was observed that for seismic denoising the number of epochs had to be significantly reduced in comparison to the initial N2V applications, whilst the number of active pixels had to be increased. The reduction of epochs hinders the network from learning to replicate mildly correlated noise whilst still providing adequate training time to learn the dominant signals in the data. Whereas, increasing the number of active pixels acts as a regulariser to the training procedure by introducing additional corruption into the training dataset.

Typically in seismic applications, denoising is performed either prior to interpretation or as preparation for down-the-line tasks such as inversion. In this paper, we took a backseat approach and choose a hyper-parameter selection that was a compromise between the three performance metrics. This resulted in a PSNR gain of 39.28\% in the image domain and 1.32\% in the inversion domain for the realistic synthetic example (Hess VTI model with bandpass filtered noise). However, if the denoising was being performed on data only for direct interpretation, we could have selected the best hyper-parameters for this task, which would have resulted in an image domain PSNR gain of 50.24\%. Similar can be said for inversion, where the inversion PSNR gain would have been 6.27\% for the optimum hyper-parameter combination (as illustrated in Figure \ref{fig:gridsearch}). Typically DL procedures are accompanied by a lengthy training time, often rendering the approaches significantly more computationally expensive than conventional procedures \cite{birnie2021arxiv}. However, due to the small number of epochs required, the N2V approach can be trained and subsequently applied within a matter of minutes for the field data - $7$ minutes for our field data experiment training. Where post-stack volumes are available, an extension to 3D denoising would be possible through the adaption from 2D convolutional layers in the NN to 3D convolutional blocks. This would likely further improve the denoising procedure at the price of increasing the computational cost and memory requirements of the network. 

The potential of N2V was illustrated using post-stack seismic data, however there is no limitation on the processing stage at which blind-spot networks could be applied for denoising. The post-stack scenario was used due to the availability of a field dataset that is known to be contaminated by random noise and that has been extensively investigated by others as a benchmark dataset for random noise suppression procedures (e.g., \cite{liu2013,liu2018}. However, in theory the technique could equally be applied to shot-gathers, receiver gathers, or even passive seismic data, assuming each of these are contaminated by random noise. One known limitation of N2V is the assumption of statistical independence between samples. \cite{broaddus2020} proposed an extension to the N2V workflow to adapt the approach for structured noise suppression. Structured N2V utilises selective masking to minimise any contribution of correlated noise into the prediction of the active pixels value. This extension suggests the potential of utilising self-supervised networks for the suppression of correlated noise signals in seismic data. 

Finally, in the initial publication of N2V, the authors acknowledge that: ``Intuitively, N2V cannot be expected to outperform methods that have more information available during training.". In other words, N2V is likely to be outmatched by well trained, supervised denoising networks. However, as noted above, creating seismic datasets with the noisy-clean pairs required for training traditional supervised procedures is not trivial. When noisy-clean pairs are generated using a previous denoising technique, such as by the Curvelet transform, the inclusion of DL can only serve to speed up the original denoising procedure. As the network learns from the training samples provided then it cannot outperform the denoising technique which was used to generate the training data. Alternatively, when synthetic data is generated to act as the training dataset, the clean image will definitely not contain any noise residual. However, generating synthetic data that accurately represent field data is a well-known challenge \cite{birnie2020}. Therefore, whilst certain steps can be taken to reduce the synthetic-field data gap for DL applications \cite{alkhalifah2021}, there is no guarantee that a synthetically trained network will be as effective when applied to field data. Recently, \cite{laine2019} proposed the first blind-spot network procedure that was shown to perform on-par with, and sometimes outperform, supervised denoising approaches for natural images contaminated by independent and identically distributed additive Gaussian noise. Future studies will consider circumstances under which self-supervised networks, of varying architectures and training procedures, can outperform supervised networks trained on synthetic seismic data, and vice versa.

\section{Conclusion}
We have shown how blind-spot networks can be applied to accurately predict seismic signals without replicating noise, and as such, provide a powerful random noise suppression procedure. As a self-learning procedure, no additional data are required for training, removing the common barriers of most deep learning denoising procedures that often require a `clean' training dataset. The Noise2Void method has been successfully applied on two synthetic and one field datasets. Whilst originally developed for random, additive, white noise, our numerical results show that such networks can be successfully trained to also remove partially correlated noise provided that the number of training iterations is reduced whilst the number of corrupted pixels is increased.

\section{Acknowledgements}
The authors thank A. Krull, T.-O. Buchholz, and F. Jug for open-sourcing their TensorFlow implementation of Noise2Void. For computer time, this research used the resources of the Supercomputing Laboratory at King Abdullah University of Science \& Technology (KAUST) in Thuwal, Saudi Arabia.

\newpage

\bibliographystyle{unsrt}  
\bibliography{bibliography}

\clearpage
\section*{Appendix: A statistical interpretation of blind-spot networks}

In this Appendix, we provide a statistical interpretation of the blind-spot networks used in this work following a derivation similar to that of \cite{laine2019} and \cite{batson2019}.

First of all, we recall the Maximum Likelihood Estimator (MLE) that is generally used as the starting point for the derivation of supervised learning training strategies:
\begin{equation}
\label{eq:mle}
\hat{\theta} = \underset{\theta}  {\mathrm{argmax}} \; p(\mathop{\mathbb{Y}}|\mathop{\mathbb{X}},;\theta)
\end{equation}
where $\mathop{\mathbb{X}}$ and $\mathop{\mathbb{Y}}$ are the distributions of the input and target data, respectively. Such distributions are generally unknown, but a set of $(x_i, y_i)\;i=1,2,..,N_s$ samples are available. Under the assumption that such samples are drawn independently from the underlying distributions, the MLE can expressed as:
\begin{equation}
\label{eq:mle1}
\begin{split}
\hat{\theta} & = \underset{\theta}  {\mathrm{argmax}} \; \prod_i p(y_i|x_i;\theta) \\
& = \underset{\theta}  {\mathrm{argmax}} \; \sum_i log(p(y_i|x_i;\theta)) \\
& \approx \underset{\theta}  {\mathrm{argmax}} \; \mathop{\mathbb{E}_{x,y \sim \mathop{\mathbb{X}}, \mathop{\mathbb{Y}}}} [log(p(y|x;\theta))],
\end{split}
\end{equation}
where the trainable parameters $\theta$ are obtained by maximizing the mean of the log-likelihood evaluated over all available pairs of inputs and targets.

In the context of denoising, the noisy image $y$ is expressed as the clean images $x$ corrupted by some noise $n$ with possibly known statistical properties i.e., $y=x+n$. However, as discussed in the main text, availability of clean images is not always possible: therefore, blind-spot networks assume that the unknown clean values depend on the context of neighboring (noisy) pixels denoted as $\Omega_x$. Moreover, when the noise can be assumed to be uncorrelated from pixel to pixel, the noisy image is used as target under the assumption that the network will only be able to reproduce its coherent part and fail to recreate the noise component, i.e. $y \approx f_\theta(\Omega_x)$. In mathematical terms the MLE can be rewritten as:

\begin{equation}
\label{eq:mle2}
\begin{split}
\hat{\theta} & = \underset{\theta}  {\mathrm{argmax}} \; \mathop{\mathbb{E}_{x \sim X}} [log(p(x|\Omega_x;\theta))] \\
& = \underset{\theta}  {\mathrm{argmin}} \; - \frac{1}{N_s} \sum_i log(p(x_i|\Omega_{x_i};\theta)), \\
\end{split}
\end{equation}
where we consider here for simplicity the $i-th$ training patch. Summing over all available patches leads to the loss function in equation \ref{eq:aetraining}.

\begin{figure*}[!htb]
  \centering
  \includegraphics[width=0.75\textwidth]{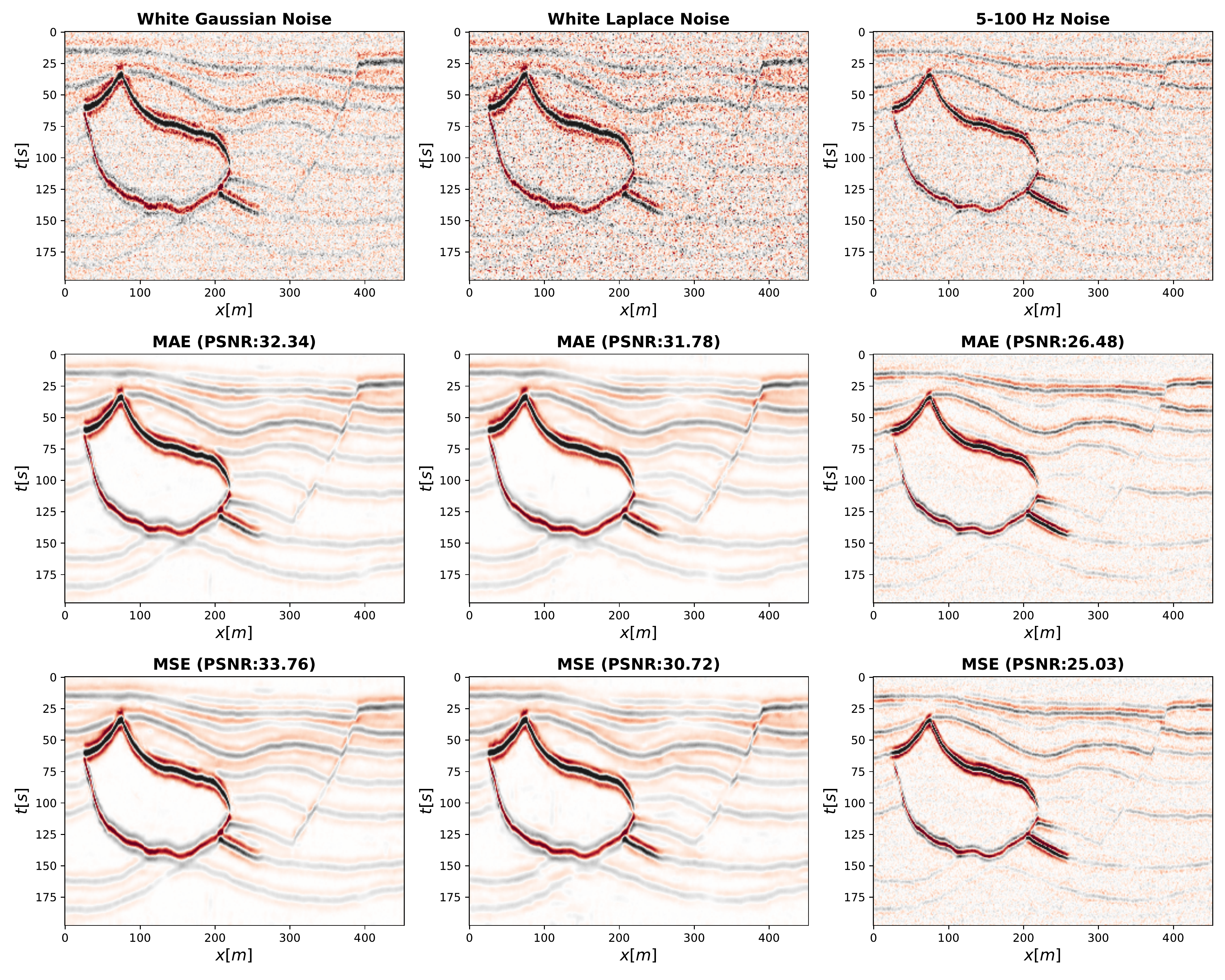}
  \caption{Denoising of synthetic dataset for 3 different noise models: left) White Gaussian noise, center) Laplace noise, right) Band-passed Gaussian noise. Top) Noisy data, middle) Denoised data using MAE as training loss, bottom) Denoised data using MSE as training loss.}
  \label{fig:ap_losses}
\end{figure*}

Let's now consider the two most commonly used statistical distributions for the noise field $n$ in seismic data and identify the corresponding MLE estimator:

\begin{itemize}
  \item \textbf{White Gaussian noise:} $n \sim \mathcal{N}(0, \sigma^2)$. The corresponding noisy image is distributed as $x\approx f_\theta(\Omega_x) - n \sim \mathcal{N}(f_\theta(\Omega_x), \sigma^2)$. Given the probability density function:
\begin{equation}
\label{eq:wgn}
p(x|\Omega_x;\theta) = \frac{1}{\sigma\sqrt{2\pi}} e^{-\frac{1}{2 \sigma^2} \left( x - f_\theta(\Omega_x) \right)^2}
\end{equation}
its corresponding training loss function becomes the well-know Mean Square Error (i.e., p=2 in equation \ref{eq:aetraining}):
\begin{equation}
\label{eq:wgnnll}
- \sum_i log(p(x_i|\Omega_{x_i};\theta)) = \sum_i \left(x_i - f_\theta(\Omega_{x_i}) \right)^2
\end{equation}
 \item \textbf{Laplace noise:} $n \sim \mathcal{L}(0, \sigma)$. The corresponding noisy image is distributed as $x \approx f_\theta(\Omega_x) - n \sim \mathcal{L}(f_\theta(\Omega_x), \sigma)$. Given the probability density function:
\begin{equation}
\label{eq:lap}
p(x|\Omega_x;\theta) = \frac{1}{2\sigma} e^{-\frac{|x-f_\theta(\Omega_x)|}{\sigma}}
\end{equation}
its corresponding training loss function becomes the well-know Mean Absolute Error (i.e., p=1 in equation \ref{eq:aetraining}):
\begin{equation}
\label{eq:lapnll}
- \sum_i log(p(x_i|\Omega_{x_i};\theta)) = \sum_i |x_i - f_\theta(\Omega_{x_i})|
\end{equation}
\end{itemize}
A numerical validation of the correspondence between statistical models for the additive noise and the choice of the training loss function is finally provided in Figure \ref{fig:ap_losses}. MSE and MAE provide the best denoising performance in terms of PSNR for Gaussian and Laplace noise, respectively.

Finally, as noise in seismic data is generally correlated in time, space, or both, we observe that neither of the above defined models is correct. On the other hand, if we assume the noise to have a certain correlation length in either time and/or space, we can express the noise within the correlation window as $\textbf{n} \sim \mathcal{N}(\textbf{0}, \boldsymbol\Sigma)$ where $\boldsymbol\Sigma$ is the covariance matrix of the noise. To take into account such correlation we must therefore write $\textbf{x} \approx f_\theta(\boldsymbol\Omega_\textbf{x}) - \textbf{n} \sim \mathcal{N} (f_\theta(\boldsymbol\Omega_\textbf{x}), \boldsymbol\Sigma)$ where we have grouped the nearby correlated pixels to form the vectors $\textbf{x}$ and $\textbf{n}$. The corresponding probability density function becomes:

\begin{equation}
\label{eq:corr}
p(\textbf{x}|\boldsymbol\Omega_\textbf{x};\theta) = \frac{1}{(2\pi)^{k/2}det(\boldsymbol\Sigma)^{1/2}} e^{-\frac{1}{2} \left( \textbf{x} - f_\theta(\boldsymbol\Omega_\textbf{x}) \right) \boldsymbol\Sigma^{-1} \left( \textbf{x} - f_\theta(\boldsymbol\Omega_\textbf{x}) \right)},
\end{equation}
and the training loss can be written as:
\begin{equation}
\label{eq:corrll}
- \sum_i log(p(\textbf{x}_i| \boldsymbol\Omega_{\textbf{x}_i};\theta)) = \sum_i \left( \textbf{x}_i - f_\theta(\boldsymbol\Omega_{\textbf{x}_i}) \right) \boldsymbol\Sigma^{-1} \left( \textbf{x}_i - f_\theta(\boldsymbol\Omega_{\textbf{x}_i}) \right),
\end{equation}

We observe that if the covariance matrix is unknown, neither MAE nor MSE can correctly approximate this loss function. In this case, empirical evidence in Figure \ref{fig:ap_losses} supports our choice of using MAE in the case of mildly correlated noise, although more sophisticated denosing models that take into account noise correlation will be investigated in the context of seismic data denoising.

\end{document}